\documentclass[conference]{IEEEtran}
\IEEEoverridecommandlockouts

\usepackage{cite}
\usepackage{amsmath,amssymb,amsfonts}
\usepackage{graphicx}
\usepackage{textcomp}
\usepackage{xcolor}
\usepackage{booktabs}
\usepackage{enumitem}
\usepackage{multirow}
\usepackage{array}
\usepackage[table]{xcolor}
\usepackage{tabularx}
\usepackage{url}

\def\BibTeX{{\rm B\kern-.05em{\sc i\kern-.025em b}\kern-.08em
    T\kern-.1667em\lower.7ex\hbox{E}\kern-.125emX}}

\begin{document}

\title{SoundscapeAgent: Agentic Soundscape Construction for Controllable Synthesis and Scalable Audio-Language Supervision}

\author{
\IEEEauthorblockN{
Hao Zhang\textsuperscript{1,*,$\dagger$},
Yiwen Zhao\textsuperscript{2,*},
Yixuan Zhang\textsuperscript{2},
Yiwen Shao\textsuperscript{2},
Steve Yves\textsuperscript{2}
}

\IEEEauthorblockA{
\textsuperscript{1}Wuhan University, Wuhan, China\\
\textsuperscript{2}Tencent Hunyuan, Bellevue, WA, USA
}

\thanks{* Equal contribution. Work done during Yiwen Zhao's internship at Tencent.}
\thanks{$\dagger$ Corresponding author: Hao Zhang (h.zhangnwpu@gmail.com).}
}


\maketitle
\begin{abstract}
We present an agentic soundscape construction framework for controllable compositional audio generation that makes explicit the scene planning, source selection, temporal layout, and rendering steps typically handled implicitly by single-shot text-to-audio models. An LLM-based agent converts user intent into an executable scene plan, acquires assets through retrieval and on-demand generation, renders controllable multi-event mixtures, and exports aligned scene metadata. The framework also supports human-in-the-loop interaction through user-guided tool selection and editable scene plans. Together, these components provide an inspectable and reusable approach to controllable soundscape synthesis and scalable audio-language data construction. Listener studies and objective metrics demonstrate competitive generation performance against text-to-audio baselines, while models trained with agent-generated data consistently outperform real-only baselines in downstream audio reasoning. Code, demos, and listening-test results are available at \url{https://haozhang6720.github.io/SoundscapeAgentDemoPage/}.

\end{abstract}


\section{Introduction}
\label{sec:intro}

Recent progress in audio-language modeling has broadened the scope of audio understanding from conventional classification and tagging to richer semantic tasks such as audio reasoning, retrieval, and event-centric interpretation \cite{ghosh2024mmau,elizalde2023clap,mei2024wavcaps,yang2025qwen3,comanici2025gemini,wu2026audio}. At the same time, growing interest in controllable and goal-directed audio synthesis has highlighted a shared underlying challenge: acoustic scenes are compositional, temporal, and multi-source, yet most generation systems treat them as undifferentiated prompt-to-waveform mappings.

Building such scenes requires coordinating scene planning, source selection, temporal layout, and rendering. However, this compositional structure is challenging to annotate manually and difficult to scale. A short audio clip may contain overlapping sources, persistent background ambience, transient foreground events, and evolving temporal relations, while coarse captions often fail to capture source roles, temporal arrangement, or event composition that downstream audio understanding may benefit from \cite{mei2024wavcaps,ghosh2024mmau}. Large-scale audio-language training therefore needs not only more paired audio-text data, but also structured supervision that makes event composition, source roles, and temporal organization explicit.

Recent advances in text-to-audio generation provide a useful starting point, with substantial progress in text-conditioned synthesis quality and emerging controllability mechanisms \cite{kreuk2022audiogen,liu2023audioldm,liu2024audioldm2,xie2024picoaudio,wang2024audiocomposer}. Nevertheless, most current systems still operate primarily as single-shot generators: scene structure is largely implicit, intermediate decisions are not exposed, and the process is difficult to inspect, edit, or repurpose for large-scale annotation generation. This limitation is increasingly visible in structured sound scene synthesis benchmarks such as the DCASE 2024 task \cite{lagrange2025dcase,task7dcase2024}.


To address these limitations, we present \textsc{SoundscapeAgent}, an agentic framework for structured soundscape construction. Building on recent LLM-driven audio systems that orchestrate external tools for compositional creation and editing \cite{huang2024audiogpt,liu2023wavjourney,liang2024wavcraft}, our framework emphasizes explicit, inspectable scene structure and aligned audio-language supervision. Rather than mapping prompts directly to waveforms, it decomposes generation into scene planning, hybrid asset acquisition, deterministic rendering, and metadata-backed description generation. The resulting pipeline supports both controllable soundscape synthesis and scalable audio-language data construction, which we evaluate along two complementary tracks.
Our contributions are:
\begin{itemize}[leftmargin=*,itemsep=1pt]
\item an agentic framework for executable scene planning, adaptive asset acquisition, layer-aware rendering, and metadata-backed export;
\item an inspectable and human-controllable design that exposes intermediate scene structure beyond single-shot synthesis;
\item a scalable corpus-construction mode that uses agent-derived scene priors to generate compositionally diverse mixtures with balanced event coverage, fine-grained temporal grounding, and aligned audio descriptions;
\item a two-track evaluation of compositional generation quality and the downstream audio-reasoning utility of agent-generated data.
\end{itemize}


\begin{figure*}[!h]
\centering
\includegraphics[width=0.99\linewidth]{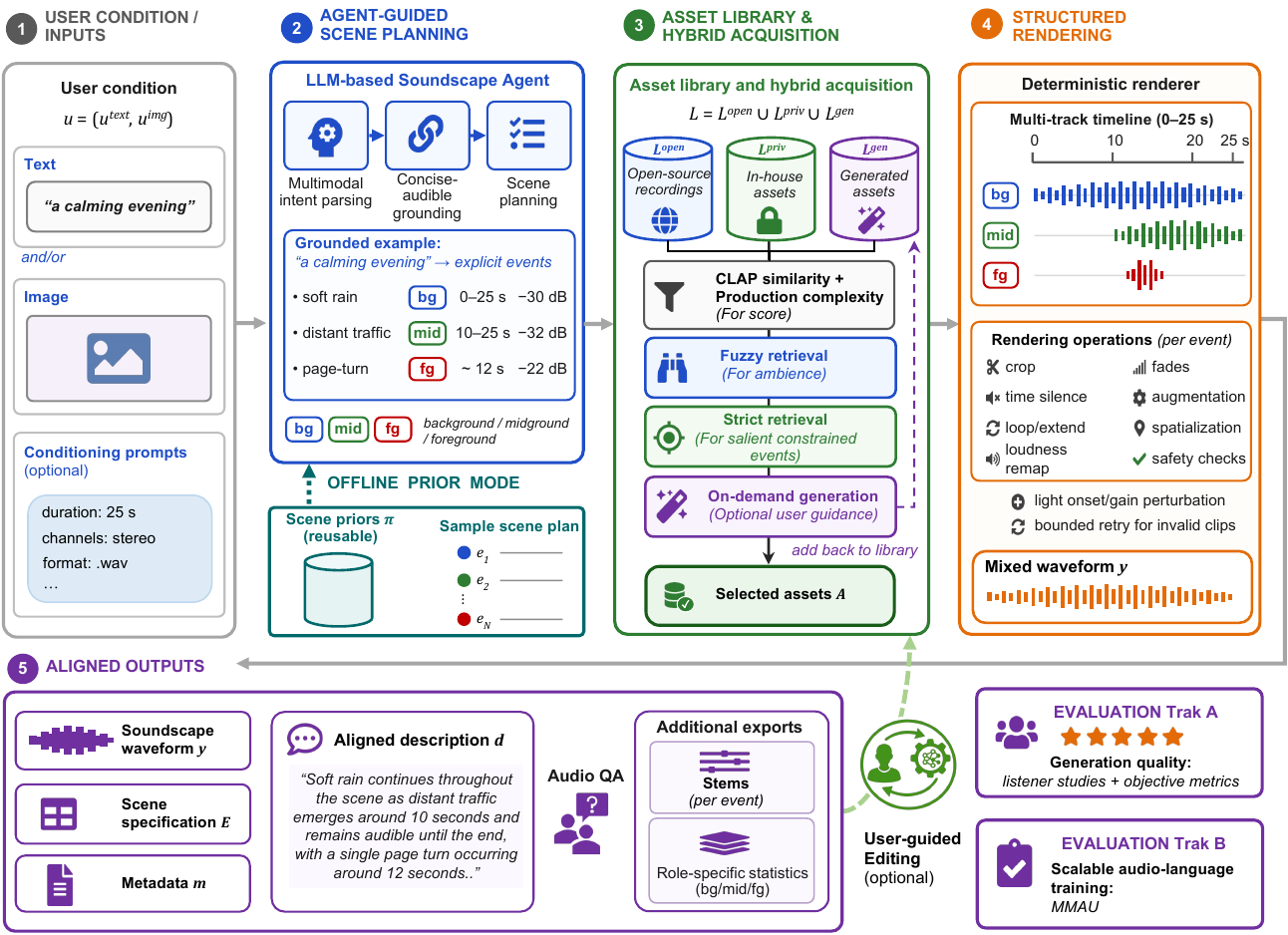}
\caption{Overview of the agentic soundscape construction framework.}
\label{fig:overview}
\end{figure*}

\section{Related Work}
\label{sec:related}

\paragraph{Audio-language supervision and audio-language models.}
Large-scale paired audio-text data has become central to modern audio-language learning. Datasets such as AudioSet \cite{gemmeke2017audioset}, AudioCaps \cite{kim2019audiocaps}, Clotho \cite{drossos2020clotho}, and WavCaps \cite{mei2024wavcaps} support audio captioning, retrieval, and downstream audio-language tasks, while contrastive language-audio pretraining (CLAP) \cite{elizalde2023clap} demonstrates the transferability of contrastive audio-text representations. Recent audio-language models extend audio understanding toward question answering, reasoning, and instruction following \cite{deshmukh2023pengi,gong2023ltu,chu2023qwenaudio,tang2023salmonn,kong2024audioflamingo}. Meanwhile, MMAU \cite{ghosh2024mmau} and recent temporal-grounding studies \cite{an2026framelevel,wang2025timeaudio} highlight the need for supervision beyond coarse captions, with richer event structure and temporal organization. Our work addresses this need through agentic soundscape construction, producing structured audio-language supervision rather than proposing a new audio-language model architecture.

\paragraph{Text-to-audio generation and controllable sound synthesis.}
Text-to-audio generation has advanced rapidly through systems such as AudioGen \cite{kreuk2022audiogen}, AudioLDM \cite{liu2023audioldm}, AudioLDM2 \cite{liu2024audioldm2}, Make-An-Audio \cite{huang2023makeanaudio}, Tango \cite{ghosal2023tango}, EzAudio \cite{hai2024ezaudio}, and TangoFlux \cite{hung2024tangoflux}. These models provide useful backbones for on-demand asset creation in our pipeline. Recent work has also explored finer-grained controllability, including timestamp and frequency control, natural-language-guided audio editing, and synchronized sound generation for video \cite{xie2024picoaudio,wang2024audiocomposer,zhang2024foleycrafter}. The DCASE 2024 sound scene synthesis task further reflects growing interest in text-controlled environmental sound generation and structured controllability evaluation \cite{task7dcase2024,lagrange2025dcase}. In contrast, our work focuses on exposing intermediate scene structure through agentic construction, enabling controllable composition, inspection, and metadata-aligned supervision beyond single-shot waveform generation.

\paragraph{Automatic data construction and agentic generation.}
Recent work has explored automatic caption enrichment and data construction to improve audio-text supervision. Sound-VECaps \cite{yuan2024soundvecaps}, for example, creates richer captions at scale and shows that stronger textual supervision can benefit text-to-audio generation. LLM-driven systems such as AudioGPT, WavJourney, and WavCraft further demonstrate that language models can orchestrate audio tools for compositional creation and editing \cite{huang2024audiogpt,liu2023wavjourney,liang2024wavcraft}. More broadly, agentic LLM studies based on reasoning-and-acting and tool-use paradigms highlight the value of agents for tasks requiring structured reasoning, tool use, and multi-step coordination \cite{yao2023react,schick2023toolformer,shen2023hugginggpt,plaat2025agentic,llmtooluse2026}. Our work follows this direction in the audio domain, but focuses on explicit scene construction: the agent exposes planning, source organization, rendering metadata, and aligned descriptions as reusable structure for controllable synthesis and reasoning-oriented supervision.

\section{Agentic Soundscape Construction Framework}
\label{sec:method}

\subsection{Overview}
\label{subsec}
We formulate compositional soundscape generation as an agent-guided construction problem, as summarized in Fig.~\ref{fig:overview} . Given a user condition $u = (u^{\text{text}}, u^{\text{img}})$, where either modality may be absent, together with optional prompts specifying requirements such as audio duration and the number of output channels, the framework produces: (i) a soundscape waveform $y$; (ii) a structured scene specification $E$; (iii) metadata $m$ recording the rendering process and scene attributes; and (iv) a textual description $d$ aligned with the generated audio.

The framework consists of two coupled layers. The \texttt{agent} layer interprets user intent and converts it into an executable scene plan that specifies acoustic events, source roles, temporal relationships, rendering attributes, and asset-acquisition actions. It then instantiates this plan through retrieval, on-demand generation, and description construction. The \texttt{renderer} layer executes the plan through deterministic timeline scheduling, waveform rendering, post-processing, and export.

The full pipeline comprises four main stages:
\begin{enumerate}[leftmargin=*, itemsep=0pt, topsep=2pt]
\item multimodal intent parsing and executable scene planning,
\item adaptive asset acquisition through retrieval or on-demand generation,
\item structured rendering and safety-aware mixing,
\item metadata-backed export and description generation.
\end{enumerate}

This design combines flexible LLM-based planning with deterministic acoustic rendering primitives. Rather than following a fixed tool sequence or mapping prompts directly to waveforms, the agent selects acquisition and rendering operations according to the requested scene and the availability of suitable assets, while explicitly exposing the intermediate scene structure. This enables controllable composition, inspectable generation, user-guided revision, and efficient large-scale data production.

\subsection{Agent-Guided Scene Planning}
\label{subsec:agent}

At the core of the framework is a soundscape agent that constructs structured scene plans rather than generating waveforms directly. Given the user condition $u$, the agent produces a scene plan that specifies acoustic events, source roles, approximate timing, relative prominence, and global constraints such as duration and output format. The agent then performs \emph{concrete-audible grounding}, translating image-based, abstract, affective, or underspecified inputs into explicit physical sound events that can be retrieved, generated, and rendered.
Formally,
\begin{equation}
(E, A, c) = \mathcal{A}(u)
\end{equation}
where $\mathcal{A}$ denotes the agent, $A$ the selected or generated asset set, $c$ scene-level control variables.

For example, the prompt ``a calming evening'' can be grounded into events such as \emph{soft rain} (background, $0$--$25$\,s, $-30$\,dB), \emph{distant traffic} (mid-ground, $10$--$25$\,s, $-32$\,dB), and a \emph{single page-turn} (foreground, $\sim$12\,s, $-22$\,dB). 
The structured plan is expanded into event candidates
\begin{equation}
{E} = \{ {e}_i \}_{i=1}^{N} 
\end{equation}
with each event represented as
\begin{equation}
e_i = \bigl(\tau_i, r_i, t_i^{\text{start}}, d_i, s_i, \ell_i, \phi_i \bigr),
\label{eq:event}
\end{equation}
where $\tau_i$ is the event tag or source category, $r_i$ the scene role , $t_i^{\text{start}}$ the onset time, $d_i$ the duration, $s_i$ the selected asset, $\ell_i$ the relative level, and $\phi_i$ optional rendering controls such as temporal jitter, panning, stereo width, or tone shaping.
Specifically, 
\begin{equation}
r_i \in \{\texttt{bg}, \texttt{mid}, \texttt{fg}\},
\label{eq:roles}
\end{equation}
where \texttt{bg} denotes persistent background ambience, \texttt{mid} denotes sustained environmental content, and \texttt{fg} denotes salient foreground or accent events. These roles guide asset selection, event density, gain assignment, and temporal placement. For image-conditioned input, visible environments, objects, and scene context are mapped to plausible sound sources through the same planning procedure.

\subsection{Asset Library and Hybrid Acquisition}
\label{subsec:assets}

Retrieval-based construction relies on a prepared single-event asset library, 
\begin{equation}
\mathcal{L} = \mathcal{L}^{\text{open}} \cup \mathcal{L}^{\text{priv}} \cup \mathcal{L}^{\text{gen}},
\label{eq:library}
\end{equation}
where $\mathcal{L}^{\text{open}}$ contains curated open-source recordings, 
$\mathcal{L}^{\text{priv}}$ denotes private in-house data, 
and $\mathcal{L}^{\text{gen}}$ contains generated assets. 
Open-source recordings are drawn from collection \textsc{Hive} \cite{li2026hive}, which emphasizes single-event purity and reduces co-occurrence noise in weakly labeled environmental audio. Generated assets are produced using text-to-audio models such as EzAudio and TangoFlux \cite{hai2024ezaudio,hung2024tangoflux}, conditioned on a filtered vocabulary of 524 leaf event classes.\footnote{The vocabulary is constructed from the union of single-event tags from \textsc{Hive} and publicly available datasets such as AudioCaps and WaveCaps.}

All candidate assets are filtered before being added to the library. We use CLAP text--audio similarity \cite{elizalde2023clap} to assess semantic consistency and an production complexity (PC) score \cite{tjandra2025meta} to prefer isolated, single-event recordings. In the current implementation, samples are retained when $s_{\mathrm{CLAP}}>0.32$ and $s_{\mathrm{PC}}<4$, with thresholds selected on a 100-clip development set. Open-source assets provide realism and acoustic diversity, while generated assets expand vocabulary coverage and fill retrieval gaps.

Given a scene plan, the system acquires an asset for each target event using three complementary modes. \textbf{Fuzzy retrieval} is used for broad or weakly specified ambience, where multiple semantically related sounds may be acceptable. \textbf{Strict retrieval} is used for salient or semantically constrained events, where the selected asset should closely match the planned source category. \textbf{On-demand generation} is invoked when retrieval fails to provide a suitable candidate. In this mode, users can select the text-to-audio backend and optionally revise the prompt before synthesis. Candidates are filtered according to the CLAP and PC scores. Newly generated assets are subsequently added to $\mathcal{L}^{\text{gen}}$ for future reuse. 

\subsection{Rendering, Interactive Editing, and Aligned Outputs}
\label{subsec:render}

Once assets are assigned, the renderer instantiates the final event table $E$, renders each event independently, and places them on a shared timeline to get the final waveform:
\begin{equation}
y(t) = \sum_{i=1}^{N} \alpha_i \, \mathcal{R}\!\left(e_i\right)\!\left(t-{t}_i^{\text{start}}\right),
\label{eq:mix}
\end{equation}
where $\alpha_i$ is the gain factor derived from $\ell_i$, and $\mathcal{R}(\cdot)$ is the event rendering operator, which includes cropping, silence trimming, looping or extension, loudness normalization, tone shaping, fading, augmentation, and optional spatialization. 
To avoid rigid or overly mechanical scenes, onset times and gains are lightly perturbed for non-accent layers before final mixing according to (\ref{eq:mix}). The renderer exports the final mixture, intermediate stems, and role-specific submixes. 
The explicit event structure and intermediate stems further support \textbf{user-guided editing}. Users may replace or modify individual tracks, add new events, or revise the scene plan rather than regenerating the entire soundscape.

Because the framework targets both controllable generation and traceable supervision, robustness and metadata are treated as first-class outputs. Near-silent or invalid clips are replaced under a bounded retry policy before mixing, and scene-level post-processing enforces loudness and peak constraints. The exported metadata records total duration, per-event timing and source parameters, background configuration, scene-level rendering statistics, safety outcomes, and generation provenance. The final textual description is generated from both intended and realized scene structure,
\begin{equation}
m = \mathcal{M}(u,E,y), \quad d = \mathcal{D}(u,E,m),
\label{eq:description}
\end{equation}
where $\mathcal{M}$ structures the processing metadata, and $\mathcal{D}$ generates the final description using either an LLM or a fixed-format scheme. By conditioning on the explicit event structure and rendering metadata, the description is more faithful than prompt-only captioning, as it can reference the rendered events, their roles, and their temporal organization. Together, $m$ and $d$ provide a detailed account of the generated audio and can optionally be used to construct audio question--answer pairs for training frameworks that require QA-style supervision.

\subsection{Offline Prior Mode.}

To combine rich agentic control with the practical requirements of large-scale data preparation, the framework also supports an offline prior mode for large-scale corpus construction. Rather than invoking full online agent reasoning for every training example, the agent is run offline to infer reusable scene priors $\pi$. Examples are then sampled efficiently via lightweight sampling, mixing and metadata-backed description generation:
\begin{equation}
\begin{aligned}
E \sim p(E \mid \pi), \quad
y = \mathcal{R}(E), \quad \\
m = \mathcal{M}(E,y), \quad
q = \mathcal{D}(E,m).
\end{aligned}
\label{eq:scalable}
\end{equation}
This mode preserves the structural advantages of agent-guided planning while substantially improving throughput.

\section{Experiments and Results}
\label{sec:experiments}

We evaluate the proposed framework along two complementary tracks. \textbf{Track~A} assesses controllable soundscape synthesis through subjective and objective comparisons with single-shot text-to-audio baselines, covering both overall generation quality and prompt alignment across diverse scene types. \textbf{Track~B} evaluates the utility of agent-generated data for scalable audio-language training by comparing LM-based audio reasoning under real-only and synthetic-augmented training regimes. The following subsections describe the experimental setup and results for each track.

\subsection{Track A: Agent Evaluation}
\label{subsec:track_a}

\begin{table}[t]
\centering
\caption{Prompt categories for Track A evaluation.}
\label{tab:prompts}
\begin{tabularx}{\linewidth}{p{0.20\linewidth} p{0.30\linewidth} X}
\toprule
\textbf{Category} & \textbf{Capability tested} & \textbf{Example prompt} \\
\midrule
AudioCaps & Standard caption-conditioned generation & A person snoring. \\ \midrule
Ambience-heavy & Background realism, texture, naturalness & A quiet forest after rain with distant birds and soft wind. \\ \midrule
Event-rich & Multi-source composition and source organization & A city street with passing cars, footsteps, bicycle bells, and distant construction. \\ \midrule
Temporally specified & Timing control and event ordering & Rain throughout; a dog barks around 3 seconds; thunder occurs near 8 seconds. \\ \midrule
Affective/abstract & Grounding from underspecified user intent & A tense abandoned hallway at night. \\
\bottomrule
\end{tabularx}
\end{table}

\begin{table*}[!t]
\centering
\small
\setlength{\tabcolsep}{7pt}
\renewcommand{\arraystretch}{1.12}
\caption{
Generation comparison across five scene types.
Amb.: ambiance-heavy; Temp.: temporally specified; Affect.: affective/abstract.
Human alignment scores (Align.) are reported on a 5-point scale as mean $\pm$ 95\% confidence interval across prompts, while CLAP similarity is reported as the mean across prompts. Higher values indicate better performance.
}
\label{tab:global_quality}
\resizebox{0.99\textwidth}{!}{
\begin{tabular}{lc*{11}{c}}
\toprule
\multirow{2}{*}{\textbf{Method}}
& \multicolumn{2}{c}{\textbf{AudioCaps}}
& \multicolumn{2}{c}{\textbf{Amb.}}
& \multicolumn{2}{c}{\textbf{Event-rich}}
& \multicolumn{2}{c}{\textbf{Temp.}}
& \multicolumn{2}{c}{\textbf{Affect.}} 
& \multicolumn{2}{c}{\textbf{Overall}} \\
\cmidrule(lr){2-3}
\cmidrule(lr){4-6}
\cmidrule(lr){6-7}
\cmidrule(lr){8-9}
\cmidrule(lr){10-11}
\cmidrule(lr){12-13}
& Align. & CLAP
& Align. & CLAP
& Align. & CLAP
& Align. & CLAP
& Align. & CLAP 
& Align. & CLAP \\
\midrule

\multirow{1}{*}{TangoFlux~\cite{hung2024tangoflux}}
& \underline{3.63 $\pm$ 0.20} & 0.38
& \underline{3.39 $\pm$ 0.20}
& 0.33
& \textbf{3.57 $\pm$ 0.18}
& 0.46
& \underline{3.33 $\pm$ 0.21}
&0.48
& 3.13 $\pm$ 0.21
&0.27
& \underline{3.41 $\pm$ 0.09}
& 0.39\\

\multirow{1}{*}{EzAudio~\cite{hai2024ezaudio}}
& \textbf{3.77 $\pm$ 0.18} & 0.38
& 3.01 $\pm$ 0.19
& 0.23
& 3.34 $\pm$ 0.17
& 0.41
& 2.31 $\pm$ 0.17
&0.28
& 2.49 $\pm$ 0.19
&0.03
& 2.98 $\pm$ 0.10
& 0.27\\

\multirow{1}{*}{AudioLDM~2~\cite{liu2024audioldm2}}
& 3.13 $\pm$ 0.24 & 0.31
& 3.13 $\pm$ 0.20
& 0.42
& 2.69 $\pm$ 0.18
& 0.37
& 2.34 $\pm$ 0.17
&0.39
& \underline{3.24 $\pm$ 0.19}
&0.28
& 2.91 $\pm$ 0.09
& 0.35\\

\multirow{1}{*}{\textbf{SoundscapeAgent}}
& 3.58 $\pm$ 0.20 & 0.36
& \textbf{3.84 $\pm$ 0.20}
& 0.31
& \underline{3.46 $\pm$ 0.21}
& 0.39
& \textbf{3.41 $\pm$ 0.20}
&0.32
& \textbf{3.88 $\pm$ 0.17}
&0.31
& \textbf{3.63 $\pm$ 0.09}
& 0.34\\

\bottomrule
\end{tabular}
}
\end{table*}

The proposed agent is compared with three single-shot text-to-audio baselines: TangoFlux~\cite{hung2024tangoflux}, EzAudio~\cite{hai2024ezaudio}, and AudioLDM~2~\cite{liu2024audioldm2}. Because the agent and the baselines differ fundamentally in their interfaces and generation mechanisms, this evaluation is not intended as a direct comparison of standalone waveform generators. Rather, it contrasts \emph{compositional soundscape construction} with \emph{single-shot text-to-audio generation}.

Five evaluation scenarios are considered: standard caption-conditioned generation using prompts randomly sampled from the AudioCaps test set, and four compositional settings covering ambiance-heavy, event-rich, temporally specified, and affective/abstract scenes. The prompt design and representative examples are provided in Table~\ref{tab:prompts}. To keep the listening test manageable, eight prompts are used per scenario, yielding 40 prompts and 160 10-second audio samples across the four systems.
The listening study involves 22 participants, each randomly assigned a subset of the samples. The assignments are balanced to ensure that every sample receives 10 independent ratings. Listeners provide a single score on a 5-point scale reflecting alignment with the intended scenario. The criterion is adapted to each prompt category based on the ``\texttt{Capability tested}'' shown in Table~\ref{tab:prompts}. 
Detailed listening-test materials and participant-level results are available on the demo page. 


The listening-test results in Table~\ref{tab:global_quality} show that \textsc{SoundscapeAgent} achieves the highest overall alignment score, outperforming all three single-shot baselines. Its advantage is most pronounced in the affective/abstract, ambiance-heavy, and temporally specified scenarios, where it exceeds the best respective baselines by $0.64$, $0.45$, and $0.08$. These gains demonstrate the agent's ability to ground underspecified intent, construct layered and naturalistic acoustic scenes, and control event timing through explicit planning. In the event-rich scenario, \textsc{SoundscapeAgent} achieves an alignment score $0.11$ below the best-performing baseline. It also remains competitive on standard AudioCaps prompts, indicating that its compositional construction strategy maintains strong performance on relatively straightforward caption-conditioned generation.

CLAP similarity~\cite{elizalde2023clap} is also reported as a complementary indicator. It does not explicitly capture several aspects of compositional soundscape construction, including temporal ordering, layered scene structure, background realism, and plausible grounding of affective or underspecified prompts. In addition, the agent may infer contextual events and introduce concurrent or sequential layers beyond the short input caption, potentially reducing global audio--text similarity even when the resulting soundscape remains coherent and well aligned with the user’s intent. We therefore place greater emphasis on the listening-test results.


\subsection{Track B: Scalable Data and Audio Reasoning}
\label{subsec:track_b}

\begin{table*}[t]
\centering
\caption{
Downstream audio reasoning performance on MMAU \texttt{test-mini}.
The \emph{real-only} baseline is an aligner trained on the CaptionStew corpus;
\emph{+ agentic aug.} denotes training with additional augmented data.
Numbers are best-checkpoint accuracy.
}
\label{tab:scaling_performance}
\begin{tabular}{lccccccccc}
\toprule
\multirow{2}{*}{\textbf{Model}}
& \multirow{2}{*}{\textbf{Data}}
& \multirow{2}{*}{\textbf{Best iter.}}
& \multicolumn{3}{c}{\textbf{Task-wise Accuracy}}
& \multicolumn{3}{c}{\textbf{Difficulty-wise Accuracy}}
& \multirow{2}{*}{\textbf{Overall Accuracy}} \\
\cmidrule(lr){4-6}
\cmidrule(lr){7-9}
& & & \textbf{Sound} & \textbf{Music} & \textbf{Speech}
& \textbf{Easy} & \textbf{Medium} & \textbf{Hard} & \\
\midrule

Real-only
& 573k
& 250k
& 59.04\%
& 46.41\%
& 47.75\%
& 45.09\%
& 56.40\%
& 44.49\%
& 51.05\% \\
\midrule

\multirow{3}{*}{+ agentic aug.}
& +50k
& 200k
& 63.06\%
& 47.01\%
& 49.85\%
& 47.32\%
& 56.67\%
& 51.27\%
& 53.30\% \\

& +100k
& 200k
& \textbf{66.67\%}
& \textbf{51.50\%}
& 51.35\%
& 47.32\%
& \textbf{61.67\%}
& \textbf{53.39\%}
& \textbf{56.50\%} \\

& +200k
& 250k
& 64.56\%
& 49.10\%
& \textbf{52.55\%}
& \textbf{48.66\%}
& 60.37\%
& 50.42\%
& 55.40\% \\
\midrule

Best $\Delta$ vs.\ real-only
& --
& --
& \textbf{7.63\%}
& 5.09\%
& 4.80\%
& 3.57\%
& 5.27\%
& \textbf{8.90\%}
& 5.45\% \\
\bottomrule
\end{tabular}
\end{table*}

\begin{figure*}[!t]
\centering
\includegraphics[width=0.98\linewidth]{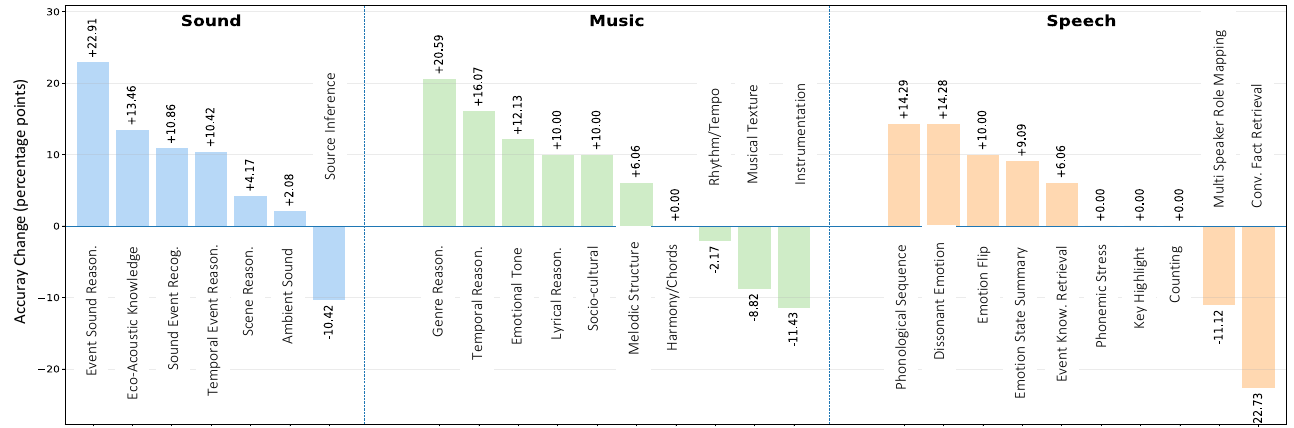}
\caption{MMAU test-mini: Sub-task accuracy change with agentic augmentation.}
\label{fig:subtasks}
\end{figure*}

This experiment addresses a central question: \emph{Does agent-structured synthetic data improve LM-based audio reasoning over real-only training?} Using the offline prior mode in Eq.~\ref{eq:scalable}, we construct a pool of 200k agent-generated examples and follow the instruction-free alignment-only recipe of~\cite{zhou2026alignment}.

The downstream model consists of a frozen Qwen2.5-Omni audio encoder~\cite{xu2025qwen25omni}, a trainable two-layer MLP projector, and a frozen Qwen2.5-7B-Instruct backbone~\cite{yang2025qwen25}. Only the projector, containing approximately 38.5M trainable parameters, is optimized. For each training clip $x$ , its description is expanded by the frozen LLM into a free-form response $y$ without system prompts or task instructions. The projector is then trained using a causal language-modeling loss on the response tokens:
\begin{equation}
\mathcal{L}(\theta) = -\mathbb{E}_{(x,y)\sim\mathcal{D}}
\sum_{t=1}^{n} \log p_L\!\left(y_t \mid y_{<t},\, P_\theta(E(x))\right),
\end{equation}
where $E$ is the frozen encoder, $P_\theta$ the projector, and $p_L$ the frozen LLM. 

Audio is represented using Whisper-v3 filter-bank features, and training clips range from 0.5 to 30 seconds. Optimization follows~\cite{zhou2026alignment} using AdamW with a peak learning rate of $10^{-3}$ and cosine decay. All experiments are conducted with \texttt{bf16} mixed precision.

All conditions use the same real-data mixture, consisting of CaptionStew-400k~\cite{tseng2025revisiting} and 173k paralinguistic examples collected in~\cite{huo2026auden}. The \textbf{real-only} baseline is trained on this mixture alone, while the augmented conditions add nested subsets of \textbf{50k}, \textbf{100k}, or \textbf{200k} agent-generated examples. The architecture, optimizer, and real-data mixture are held fixed, so only the amount of synthetic data varies.

Training is capped at 300k iterations for all conditions. Checkpoints are evaluated at 200k, 250k, and 300k iterations, and the checkpoint with the highest MMAU \texttt{test-mini} accuracy is reported.

Evaluation is conducted on MMAU \texttt{test-mini}~\cite{ghosh2024mmau}, which contains 1{,}000 audio question-answering examples spanning Sound, Speech, and Music. Decoding uses beam search with four beams and a maximum of 256 generated tokens. Predictions are scored using the official MMAU string-matching protocol. We report overall accuracy together with task-wise and difficulty-wise breakdowns.

As shown in Table~\ref{tab:scaling_performance}, augmenting the real-only training set with agent-generated data improves performance across all task and difficulty categories. The best setting, using 100k synthetic examples, raises overall accuracy from 51.05\% to 56.50\%. The largest task-wise gain occurs on Sound (+7.63 points), consistent with the sound-centered composition of the synthetic corpus, while gains on Music and Speech indicate transfer beyond the primary target domain. The largest difficulty-wise improvement is observed on Hard questions (+8.90 points), suggesting that supervision involving layered events, temporal relationships, and relative source prominence is particularly useful for complex audio reasoning. Performance does not increase monotonically with data volume, as the 200k setting underperforms the 100k setting overall. This suggests that augmentation effectiveness depends on data quality, diversity, and the balance between real and synthetic examples, rather than volume alone.

The subtask-level analysis in Fig.~\ref{fig:subtasks} further shows that the gains are concentrated in capabilities aligned with the structured properties of the generated data. On the Sound track, the largest improvements occur for Event-Based Sound Reasoning (+22.91 points), Eco-Acoustic Knowledge (+13.46), Sound-Based Event Recognition (+10.86), and Temporal Event Reasoning (+10.42), reflecting the corpus emphasis on event composition, environmental context, and temporal organization. Transfer is also observed for higher-level Music capabilities, including Musical Genre Reasoning (+20.59), Temporal Reasoning (+16.07), and Emotional Tone Interpretation (+12.13). Results on Speech are more mixed, as the asset library in Sec.~\ref{subsec:assets} primarily contains non-speech events. 
Overall, agent-generated supervision primarily benefits compositional, temporal, and contextual audio reasoning, with less consistent transfer to speech-specific and fine-grained dialogue understanding.

\section{Discussion and Conclusion}
\label{sec:conclusion}

This work treats compositional soundscape generation as an agentic construction problem rather than a single-shot mapping from text to waveform. By exposing planning, hybrid asset acquisition, and rendering as inspectable stages, the proposed framework supports both direct generation evaluation and scalable audio-language supervision. The two-track evaluation separates \emph{whether compositional agent construction produces competitive multi-event soundscapes} from \emph{whether its structured outputs provide useful training signals}.

Track~A compares compositional construction with monolithic text-to-audio baselines and evaluates controllability and grounding capabilities that are not well captured by conventional objective metrics. Track~B examines whether agent-derived scene priors provide synthetic supervision that improves LM-based audio reasoning beyond real-only training.

Current limitations include dependence on asset inventory and generator quality, the domain gap between synthetic and real audio, and evaluation coverage constrained by the selected prompts and benchmarks. In particular, some of the fine-grained information captured by our data is not explicitly evaluated by MMAU, so the current results may reflect only part of the value of the generated supervision. In addition, the present asset library is dominated by non-speech sound events. Future work will explore additional benchmarks and diagnostic tasks that better capture these structured audio-reasoning capabilities, while expanding speech and music coverage to support a broader range of downstream applications. Overall, the framework provides a reusable design pattern for controllable soundscape synthesis and a practical approach to scaling structured audio-language training data.

\section*{Ai Generated Content Disclosure}
ChatGPT (OpenAI) was used for language polishing and rephrasing in the Introduction, Method, and Experiments sections. It was not used to generate research results, conduct experiments, or draw scientific conclusions. The authors reviewed and revised all AI-assisted text and take full responsibility for the final content.

\bibliographystyle{IEEEtran}
\bibliography{custom}

\end{document}